\documentclass[%
nofootinbib,
reprint,twocolumn,amsmath,amssymb,superscriptaddress,aps,prd
]{revtex4-2}

\usepackage{graphicx}% Include figure files
\usepackage{dcolumn}% Align table columns on decimal point
\usepackage{bm}% bold math
\usepackage[colorlinks=true]{hyperref}% add hypertext capabilities
\usepackage[mathlines]{lineno}% Enable numbering of text and display math
\usepackage{aas_macros}
\usepackage{color}
\usepackage{bbm}
\usepackage{multirow}
\usepackage{subcaption}
\usepackage{siunitx}
   
\newcommand{\chisq}{\ensuremath{\chi^2}}

\newcommand{\lcdm}{$\Lambda$CDM}

\newcommand{\Omm}{\ensuremath{\Omega_\text{m}}}
\newcommand{\Omo}{\ensuremath{\Omega_{\text{m},0}}}

\newcommand{\rd}{\ensuremath{r_\text{d}}}

\newcommand{\fseight}{\ensuremath {f\sigma_8}}

\newcommand{\seighto}{\ensuremath {\sigma_{8,0}}}
\newcommand{\Geff}{\ensuremath {G_\mathrm{eff}}}
\newcommand{\fde}{\ensuremath {f_\mathrm{DE}}}

\def\tpdf#1{\texorpdfstring{#1}{Lg}}

\definecolor{deepmagenta}{rgb}{0.8, 0.0, 0.8}
\definecolor{ballblue}{rgb}{0.13, 0.67, 0.8}

\definecolor{RedWine}{rgb}{0.743,0,0}

\begin{document}

% \preprint{}

\title{Joint reconstructions of growth and expansion histories from Stage-IV surveys with minimal assumptions I: Dark Energy beyond \tpdf{$\Lambda$}}% Force line breaks with \\

%\author{Rodrigo~Calderón}\orcid{0000-0002-8215-7292}}
\author{Rodrigo~Calderón}%\,\orcidlink{0000-0002-8215-7292}}
\email{calderon@kasi.re.kr}
\affiliation{Korea Astronomy and Space Science Institute,  Daejeon 34055, Korea}
\affiliation{Laboratoire Charles Coulomb, Universit\'e de Montpellier \& CNRS, 34095 Montpellier, France}

\author{Benjamin~L'Huillier}%\,\orcidlink{0000-0003-2934-6243}}
\email{benjamin@sejong.ac.kr}
 \affiliation{Department of Physics and Astronomy, Sejong University, Seoul 05006, Korea
}%
\affiliation{Department of Astronomy \& Space Science, Yonsei University,  Seoul 03722, Korea}

\author{David~Polarski}%\,\orcidlink{0000-0002-7049-8276}}
\email{david.polarski@umontpellier.fr}
\affiliation{Laboratoire Charles Coulomb, Universit\'e de Montpellier \& CNRS, 34095 Montpellier, France}

\author{Arman Shafieloo}%\,\orcidlink{0000-0001-6815-0337}}
\email{shafieloo@kasi.re.kr}
\affiliation{Korea Astronomy and Space Science Institute, 
 Daejeon 34055, Korea}
 \affiliation{University of Science and Technology, Daejeon 34113, Korea}

\author{Alexei~A.~Starobinsky}
\email{alstar@landau.ac.ru}
\affiliation{L. D. Landau Institute for Theoretical Physics RAS, Moscow 119334, Russia}

\date{\today}

\begin{abstract} 
Combining Supernovae, Baryon Acoustic Oscillations and Redshift-Space Distortions data from the next generation of (Stage-IV) cosmological surveys, we aim to reconstruct the expansion history up to large redshifts using forward-modeling of $\fde(z) = \rho_\mathrm{DE}(z)/\rho_\mathrm{DE,0}$ with Gaussian processes (GP).
In order to reconstruct cosmological quantities at high redshifts where few or no data are available, we adopt a new approach to GP which enforces the following minimal assumptions: a) Our cosmology corresponds to a flat Friedman-Lema\^ itre-Robertson-Walker (FLRW) universe; b) An Einstein de Sitter (EdS) universe is obtained on large redshifts. 
This allows us to reconstruct the perturbations growth history from the \emph{reconstructed} background expansion history. Assuming various DE models, we show the ability of our reconstruction method to differentiate them from \lcdm\ at $\gtrsim2\sigma$.

\end{abstract}

\maketitle

\section{Introduction}

Despite its success, the concordance model of cosmology, the \lcdm\ model, is facing numerous challenges \cite{Weinberg:1988cp,Sahni:1999gb,doi:10.1146/annurev-astro-091916-055313,perivolaropoulos2021challenges}. 
On the theory side, the nature of dark energy (DE) and dark matter remain a mystery though their phenomenology is constrained from the data with increasing accuracy. 
In addition, while general relativity (GR) has been very successful on solar system scales, alternative gravity models with observable signatures on large cosmic scales remains a challenging possibility.    
Observationally, despite all the successful predictions of the concordance model, the arrival of data of ever increasing quantity and quality has led to the rise of tensions, most notably, the $H_0$ and the $S_8$ tensions \cite{riess2022comprehensive,Cosmo_Intertwined_II_2021,Cosmo_Intertwined_III_2021,DiVal2021}. 
This is another incentive to question the nature of gravity itself \cite{MG2012}.  
As it is well-known, combining background and perturbations can be a decisive tool to constrain gravity and hence the nature of dark energy \cite{Starobinsky:1998fr}. Very generally, 
extraction of physical information from the data in a way that is as reliable as possible remains an important challenge.  

To face these issues, two main approaches have been adopted by the community.   
A common approach is to confront new theories to the data, hence constraining the model parameters. 
On the other hand, one can perform a bottom-up approach from the data to the phenomenology \cite[\emph{e.g.}][]{2006MNRAS.366.1081S,2007MNRAS.380.1573S,Alam:2008at,2010PhRvD..81h3537S,2012PhRvD..85l3530S,2013PhRvD..87b3520S,2017JCAP...01..015L,2018MNRAS.476.3263L,2020MNRAS.494..819L,Raveri2021dbu,2021arXiv210704343A,2021arXiv211011421N}. 
This latter approach aims to retrieve the ``correct'' phenomenology from the data as independently as possible of theoretical assumptions.  
In this work, we take a hybrid approach, in-between traditional model-fitting and model-independent approaches.
At the core of the work presented here lies the reconstruction of the expansion history from the data with Gaussian processes (GP). 
Indeed Gaussian processes are a very effective tool when one follows an approach where no or little theoretical assumptions are made on the possible DE phenomenologies. However this method can lead to large and undesirable uncertainties at large redshifts, from $z\gtrsim 1-2$ on \cite{2020MNRAS.494..819L}. 
 This is a consequence of the method itself when only a small amount of data are available. 
 For example one should remember that  no SNIa data are available when one comes close to $z\sim 2$. 
 Nevertheless sticking to this reconstruction method in view of its advantages and elegance, we first devise a way to get reasonable uncertainties on very large redshifts: we simply impose that our universe should be close to an Einstein de Sitter (EdS) universe on large redshifts (we have in mind here redshifts still obeying $z\ll z_{\rm eq}$). 
 We assume a flat FLRW universe, where for $z\ll z_{\mathrm {eq}}$, the dimensionless expansion rate $h(z)\equiv H(z)/H_0$ is given by
\begin{equation}\label{eq:background}
    h^2(z)=\Omega_{\mathrm{m,0}}\;(1+z)^{3}+(1-\Omega_{\mathrm{m,0}})\;f_{\mathrm{DE}}\,(z)\;, 
\end{equation} 
where $f_{\mathrm {DE}}(z)\equiv\frac{\rho_{\mathrm {DE}}(z)}{\rho_{\mathrm {DE,0}}}$ encodes the evolution of the dark energy density. Equality \eqref{eq:background} covers many physical situations. As we have in mind a joint reconstruction of both the background expansion and the perturbations, we postpone a discussion of the scope and limitations of 
\eqref{eq:background}, and generally of the method presented here, after we address the evolution of the perturbations (see Eqs. \eqref{eq:growth} and \eqref{eq:growth_f} below).   
Hereafter,  EdS universe will refer to the limit when the second term of \eqref{eq:background} becomes negligibly small. 
While this is obvious theoretically, its implementation can be problematic when using Gaussian Processes.
We will reconstruct $f_{\mathrm {DE}}(z)$ using GP with reasonable minimal assumptions. In this way only these expansion histories are reconstructed for which the second term in \eqref{eq:background} is negligible compared to the dust-like matter term at large-$z$. While this can be seen, and indeed is, a theoretical prior, we believe it is altogether reasonable and leaves the possibility to consider many dark energy models which is the key point. In other words, this prior can still fit a very large number of dark energy models and these are the models we are interested in here. Another advantage of this approach is to render this assumption explicit rather than tacit. So \eqref{eq:background} fixes the theoretical framework in which we want to get the dark energy phenomenology and the reconstruction is such that we exclude behaviours at large redshifts which are in our opinion less interesting. 
In this work we will also assume a spatially flat universe as it is obvious from \eqref{eq:background} and we will comment in the conclusion on relaxing this assumption. In the next sections we present the method and check its successful use. Last but not least, we stick to General Relativity (GR) in this work, leaving modified gravity models for future work.  
The method and data are described in Section~\ref{sec:method}, 
Section~\ref{sec:valid} shows the validity of the method, and  Section~\ref{sec:varcosmo} is devoted to the reconstructions of other fiducial cosmologies.
We discuss our results in Section~\ref{sec:ccl}.

\section{Method and Data}
\label{sec:method}

\subsection{Method}\label{Method}
 
In this paper, we aim to reconstruct the expansion and growth histories without assumptions concerning the dark energy component and assuming GR. 
To do so, as we have said earlier, we assume a flat FLRW universe, where for $z\ll z_{\mathrm {eq}}$, the dimensionless expansion rate $h(z)\equiv H(z)/H_0$ is given by \eqref{eq:background}. 
We note that the \fde\ term is essentially free to capture a large variety of DE behaviours.  

In General Relativity, the equation for the growth of perturbations in the Newtonian approximation
valid on observable (sub-Hubble radius) cosmic scales is given by
\begin{align}
\label{eq:growth}
    \ddot\delta + 2H~\dot{\delta} & = 4\pi G~\rho ~\delta,
\end{align}
where a dot stands for a derivative with respect to time, $\delta$ is the relative energy density fluctuation of dust-like matter and $\rho$ its background energy density.  
The above equation can also be re-written in terms of the growth factor $f\equiv\delta'/\delta$, to give
\begin{align}\label{eq:growth_f}
f' + \left(f+2+\frac{h'}{h}\right) f - \frac 3 2 \Omm = 0~,
\end{align}
where a prime stands for a derivative with respect to $\ln a$. 
Let us discuss now the variety of situations for which our theoretical set-up holds.
In models where DE perturbations are negligible, we can use Eqs. \eqref{eq:growth} and \eqref{eq:growth_f} as they
stand. From this point of view, the two essentially dark components appearing in \eqref{eq:background} are divided in the following way: only the first term includes baryonic matter and cold dark matter, which undergo gravitational clustering at the comoving scale corresponding to $8\,h^{-1}\rm Mpc$ today. The second term can even include hot non-relativistic matter, for example
massive neutrinos, for which the free-streaming scale is much larger than this scale. 
Of course, strictly speaking, while $\Omega_{\mathrm{m,0}}$ appearing in the first term of \eqref{eq:background} corresponds to 
clustered matter -- crucially, it is this quantity which would appear in the driving term of the equation for the growth of matter perturbations \eqref{eq:growth} and \eqref{eq:growth_f} -- the parameter $\Omega_{\mathrm{DE,0}}$ on the other hand includes all remaining unclustered components besides DE, and the same applies to $f_{\mathrm{DE}}$. But these effects remain tiny because 
$\Omega_{\nu} \ll 1$. In the same way we do not include here the effect of massive neutrinos on the growth of perturbations which translates into a very tiny damping during the matter era with $\delta_m\propto a^{1-\frac35 \Omega_{\nu}}$ for $\Omega_{\nu} \ll 1$. 
This cumulative effect becomes however relevant if we integrate back from very high redshifts of order $z_{\mathrm{eq}}$, which is not the case here. In this context, we note that the presence of any hot matter can be pinpointed by considering matter perturbations on larger comoving scales, but there are unfortunately few such accurate data available at the present time. 
 
We note further that equations 
(\ref{eq:background}--\ref{eq:growth_f})
can in principle apply to non-interacting tracking (early) dark energy. In that case DE tends to a dust-like behaviour with $\Omega_{\mathrm{DE}}\to \Omega_{\mathrm{DE},\infty}\ne 0$ 
($\Omega_m\to \Omega_{m,\infty}$), with $\Omega_{\mathrm{DE},\infty} \ll \Omega_{m,\infty}$ from observations (see e.g. \cite{Planck:2015fie}). 
We stress again that when tracking DE or massive neutrinos behaves like dust, their contribution to \eqref{eq:background} is very small compared to the first term. 

Interestingly, the second term in \eqref{eq:background} can in principle also include the energy density of dark matter-dark energy interactions in models where these exist. However this term is known to be small because observations severely restrict deviations of $w_{\mathrm{DE}}$ from $w_{\Lambda}=-1$ on one hand, but also because these interactions, whether physical or purely gravitational as in dark energy models based on scalar-tensor gravity \cite{2000PhRvL..85.2236B}, would result in a small gravitational clustering of the components represented by the second term due to clustering of the dust-like matter components, and we do not consider this clustering here.   
\smallbreak

Observationally, redshift-space distortion (RSD) provides us with growth rate measurements of the (unbiased)
quantity\footnote{We note that the pipelines used to extract $\fseight$ themselves assume a fiducial cosmology, which may introduce a bias.}
\begin{equation}% \label{eq:f}
f\sigma_8\equiv\frac{\sigma_{8,0}}{\delta_0} f\delta = \frac{\sigma_{8,0}}{\delta_0}\delta',
\;\;\;\text{ with }\;\;\;
\delta_0=\delta(z=0).
\end{equation}

To validate the method, we proceed as follows. 

\begin{figure}
    \centering
    \includegraphics[width=\columnwidth]{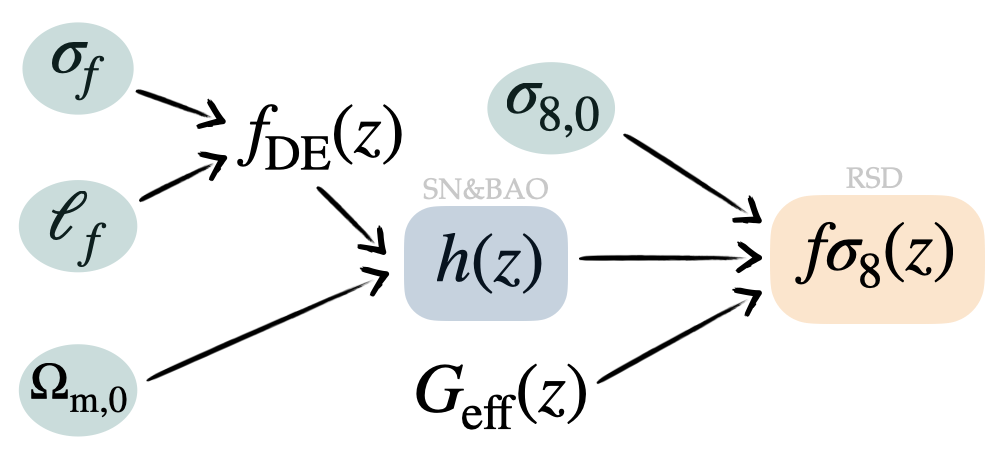}
    \caption{Diagramatic representation of our method. Given a set of hyperparameters $(\sigma_f,\ell_f)$ we obtain a behaviour of \fde, which together with a value of \Omo, yields an expansion history $h(z)$. We then solve the growth equation assuming a $\Geff$ (though in the present work we stick to GR, where $G_{\rm eff}= {\rm const} = G$) and a given $\seighto$ to obtain the growth history $\fseight$. The values for the parameters (in light-blue) are obtained through MCMC sampling.%\rcb{We might want to keep this figure or not} \ben{I think it's a very efficient visual summary of the method, so i'd want to keep it.}
    }
    \label{fig:diagram}
\end{figure}
\begin{enumerate}
\item Generate the mock data:
    \begin{enumerate}
        \item Choose a fiducial cosmology 
        \item Generate the background (SN, BAO) and perturbations (RSD) mock data according to this cosmology.
    \end{enumerate}
        \item Reconstruct the expansion and growth histories using MCMC:
         Choose the cosmological parameters  $(\Omo,\seighto)$ and the hyper parameters $(\sigma_f,\ell_f)$ via MCMC. % or by fixing them. 
        
        For each point in the parameter space: 
    \begin{enumerate}
        \item Generate a single realization of $\fde\sim\mathcal{GP}(-1,K(\sigma_f,\ell_f))$---imposing $f_{\rm DE,0}\equiv1$.  
        \item Obtain the expansion history $h(z)$ via Eq.~\eqref{eq:background} for $\fde, \Omo$ 
        \item Solve the growth equation~\eqref{eq:growth_f} and obtain the growth history $\fseight(z)$
        \item Confront the expansion and the growth histories to the mock data and obtain a likelihood for this choice of (hyper)parameters.
    \end{enumerate}
\end{enumerate}
A diagrammatic representation of our method is displayed on Figure  \ref{fig:diagram}. 
In \S~\ref{sec:valid}, we will show the effect of different choices of mock datasets on the reconstructions of the expansion history. 
% \clearpage

\subsection{Gaussian Process modelling}\label{GPR}
\begin{table}
    \centering
    \begin{tabular}{ccccc}
        \hline
		Parameter & $\Omega_{\rm m,0}$ & $\sigma_{8,0}$ & $\sigma_f$ & $\ell_f$ \\ 
		\hline
		Prior & $[0.1,0.6]$ & $[0.6,1.2]$ & $[10^{-2},2.5]$ & $[10^{-2},7]$ \\ 
		\hline
    \end{tabular}
    \caption{Uniform priors for the parameters used in the MCMC analyses.}
    \label{tab:priors_params}
\end{table}

In this study, we need to know the expansion history up to high redshift in order to solve the growth equation~\eqref{eq:growth}. 
Gaussian Processes (GP) have been extensively used in the literature \cite{2010PhRvD..82j3502H,2010PhRvL.105x1302H,2011PhRvD..84h3501H,2012PhRvD..85l3530S} to reconstruct the expansion history directly from the data. 

A clear downside of the usual application of GP is that one can only reconstruct a quantity where there are data, as in the absence of data, the GP will fall back to its mean function. 
Therefore, modeling $h$ as a GP can 
(i) lead to an incorrect asymptotic behaviour of $h$ (because of the lack of data at high-$z$)
and (ii) does not disentangle the matter term \Omo\ from the dark energy term
, which is necessary to solve the growth equation \eqref{eq:growth} in a consistent way. 
Hence, we must find a way to impose $h^2(z\gg 1) \sim \Omo  (1+z)^3$: 
we enforce this EdS 
limit regardless of the value of \Omo, by modelling \fde\ as a GP with mean $\bar{f}=1$ and imposing   $f_{\mathrm {DE}}(z=0)=1$. 

 This ensures the correspondence between the \Omo\ part of $h$ and \Omo\ appearing in the right-hand side of the growth equation \eqref{eq:growth}.
 
 We choose an exponential squared kernel as covariance function \cite{rasmussen2006gaussian} such that 
 \begin{equation}
    f_{\mathrm {DE}}(z)\sim\mathcal{GP}\left(\bar f=1,k=K(\sigma_f,\ell_f)\right)
\end{equation}
where $\sigma_f$ and $\ell_f$ are the usual hyperparameters controlling the deviations from the mean and the correlation length in dataspace, respectively. % We further impose $f_{\mathrm {DE}}(z=0)=1$. 
We can then draw samples of $f_{\mathrm {DE}}(a)$ from this multivariate (Gaussian) distribution. 
Substituting $f_{\mathrm {DE}}$ in \eqref{eq:background} provides us with a family of samples of $h(z)$ that we can use to solve for the perturbations $\delta_m(z)$ using \eqref{eq:growth} (or equivalently \eqref{eq:growth_f}) without assuming a particular DE model. 
This approach uses \emph{untrained} GP rather than \emph{trained} GP, and calculate the goodness of fit of each sample. It is similar to the approach in \cite{2018PhRvD..97l3501J} and \cite{2020MNRAS.494..819L}. 
The advantage of using \emph{untrained} GP over trained GP is that one can perform forward modeling, and therefore is not restricted to the redshift range where the data are. 
This is particularly interesting in our case: since we need to solve the growth equation \eqref{eq:growth} with the correct initial conditions. 
By modeling \fde\ as a GP with mean function one, we enforce that at high-redshift, the expansion history is EdS, and we can solve the growth equation. 
In this approach, rather than finding the best-fit hyperparameters that maximize the marginal likelihood as is commonly done\cite{rasmussen2006gaussian}, we use a more Bayesian approach and marginalize over the hyperparameters \cite{2018PhRvD..97l3501J,2020MNRAS.494..819L,2022arXiv220107025R}.
For a detailed comparison between trained and untrained GP and on the effect of marginalizing over the hyperparameters, see \cite{seunggyu}.

% \smallbreak
We perform our analysis in increasing levels of complexity. 
We first fix the cosmological parameters (\Omo,\,\seighto) to their fiducial value, and sample the (hyper)parameter space using Markov Chain Monte Carlo (MCMC) methods as implemented in \texttt{emcee}~\cite{Foreman_Mackey_2013}. 
We then include the cosmological parameters one by one in the MCMC analysis to asses their individual effect on the reconstructions. 
Table \ref{tab:priors_params} shows the prior ranges used throughout this work.
The hyperparameters $(\sigma_f,\,\ell_f)$ will determine the shape of \fde, which together with a given value of \Omo, determines the expansion history $h(z)$ through Eq.\eqref{eq:background}. 
Solving Eq. \eqref{eq:growth} and translating into $f\sigma_8(z)$ further requires the knowledge of \seighto\, and a given shape of \Geff$(k,z)$. 
Again, in this work we restrict ourselves to the case of GR.

\begin{figure*}
    \centering
    \includegraphics[width=\textwidth]{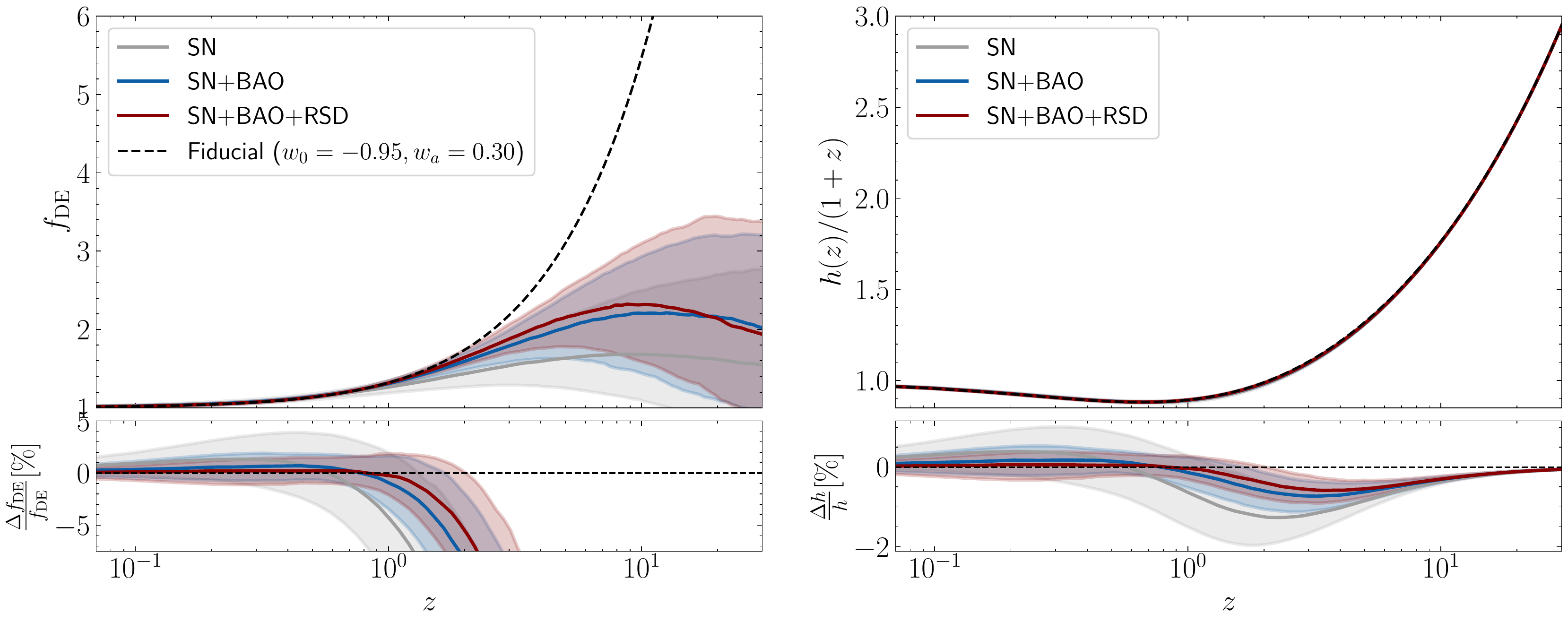}
    \caption{Left: reconstructed \fde\ from various combinations of the data, assuming perfect knowledge of $(\Omo, \seighto)$. Right: reconstructed $h(z)/(1+z)$. Solid lines and shaded regions correspond to the median and 68\% confidence intervals around it, respectively. The dashed-black line correspond to the true (fiducial) cosmology used to generate the mock data. 
    }
    \label{fig:h_om_s8}
\end{figure*}
\subsection{Data}
\label{sec:data}
\subsubsection{Mocks data }
\label{sec:mock}

To validate our method, we proceed as follows. We first generate mock data from a fiducial CPL cosmology, with $\theta^{\mathrm {fid} }=\{\Omega_{\mathrm{m,0}} ^{\mathrm{fid}}=0.28,w_0^{\mathrm {fid}}=-0.95,w_{a}^{\mathrm {fid}}=0.3,\sigma_{8,0}^{\mathrm {fid}}=0.81\}$.
In this fiducial model one has \cite{2001IJMPD..10..213C,2003PhRvL..90i1301L}
\begin{equation}\label{eq:Ode_over_Om}
\frac{ \Omega_{\rm DE} }{ \Omega_{m} } = \frac{ \Omega_{\rm DE,0} }{ \Omega_{m,0} } (1+z)^{3(w_0 + w_a)} 
\exp{\left( -3 w_a \frac{z}{1+z} \right)}~.
\end{equation}
Hence for the interesting range $-1 < w_0 + w_a < 0$, while $\Omega_{\rm DE}\to 0$ in the past, it does so more slowly than if DE was in the form of a cosmological constant $\Lambda$. 
As an illustration, we get for any given values $\Omega_{\rm DE,0}$ and $\Omega_{m,0}$ at $z=99$ (chosen for convenience)
\begin{equation}\label{eq:Ode_over_Om-fid}
\frac{ \Omega^{\rm fid}_{\rm DE} }{ \Omega_{m} }(z=99) = 
\frac{ \Omega_{\rm DE,0} }{ \Omega_{m,0} }~f^{\rm fid}_{99} \times 10^{-6}~,
\end{equation}
where we have defined 
\begin{equation}\label{f99} 
f_{99} \equiv f_{\rm DE}(z=99)~,  
\end{equation}
Using $f_{99}^{\rm fid}=51.64$, it is seen that the ratio $\frac{ \Omega^{\rm fid}_{\rm DE} }{ \Omega_{m} }(z=99)$, while very tiny, is 
still larger by a factor $51.64$ than it would be for a cosmological constant $\Lambda$. 
As we shall later discuss in Section~\ref{sec:bias}, for the particular fiducial values chosen here, this can lead to a biased determination of \Omo\ and \seighto.
We generate Mock SNIa, BAO, and RSD data as follows:
\begin{itemize}
    \item SNIa: we generate Roman-like (formerly ``WFIRST)
    data following Tab.7 in \cite{Riess:2017}. Due to the high-computational nature of the problem, we use this compressed likelihood made of {9 points} in $h(z)$. We note that these assume a flat Universe. We will refer to this dataset as ``SN''.
    \item BAO:  we simulate a DESI-like survey, following Tabs. 2.3-2.7 in \cite{2016arXiv161100036D}. More specifically, we consider the $14,000 \deg^2$ survey as baseline. 
    The BAO measurements are given in terms of the combinations $H(z) \rd$ and $d_A(z)/\rd$.
    Since we do not include any early universe information in our analysis, we express our \chisq\ in terms of the product $H_0\rd$. For the method validation with mock data, and for computational reasons, we fixed $H_0r_d$ to its fiducial value\footnote{We tested including the combination $H_0r_d$ as a free parameter, finding that a (Stage-IV) SN+BAO+RSD analysis is able to constrain the product to be within the $1\sigma$ region of its true (fiducial) value.}. %.\rcb{See Fig/"Triangle all free.png"}}. 
    In the case of (real) current data, it is left as a free parameter, as we ignore its true value.
    \item RSD: we generate \fseight\ measurements for a DESI-like survey, always following  Tabs. 2.3-2.7 in \cite{2016arXiv161100036D} from the fiducial cosmology. Note that these require assuming a given \Geff. We refer to these measurements as ``RSD''.
\end{itemize}

We note at this stage that the acoustic peaks in the CMB allow for a precise determination of the quantity $100\,\theta_s(z_\star)=100\, r_s(z_\star)/r(z_\star)=1.04110\pm0.00031$ where $r_s(z_\star)$ is the comoving sound horizon and $r(z_\star)$ is the comoving distance at recombination. 
We use this additional constraint at recombination, hence at redshifts much larger than our numerical calculations, to check that all our models have an 
acceptable behaviour at very large $z$.
Using the Planck values $z_\star=1089.92$, $r_s(z_\star)=\SI{144.4}{Mpc}$ and $H_0=\SI{70}{km.s^{-1}.Mpc^{-1}}$, 
a calculation of the quantity $100\,\theta_s$ 
for the fiducial CPL model considered here
yields 1.0427 close to the one reported by Planck \cite{Planck:2018vyg}.  
Since we deal with mock data and we do not include any early-universe information in the analysis, we 
expect that inclusion of data at high-$z$ will only confirm and refine our results. Similar values are obtained for the rest of the fiducial cosmologies considered in Section \ref{sec:varcosmo}.
\smallbreak

We performed a series of separate and joint analysis, using SN, BAO and RSD data. In the next section, we illustrate our reconstructions for a fiducial cosmology with a substantial departure from \lcdm~---depicted by the dashed line in Fig. \ref{fig:h_om_s8}--- to test a ``worst-case'' scenario.
This enables us to asses whether our framework allows for an accurate reconstruction of \fde\ and thus, of the expansion history $h(z)$.

\section{Validation of the Method with Mock Data}
\label{sec:valid}
In this section, we apply our formalism to the mock data generated above in order to validate the method in the case of GR. 
We first assume perfect knowledge of $(\Omo,\seighto)$, then we will only assume $\seighto$, and finally we do not assume any parameter to reconstruct the growth and expansion histories.
\begin{figure}%[b]
    \centering
    \includegraphics[width=\columnwidth]{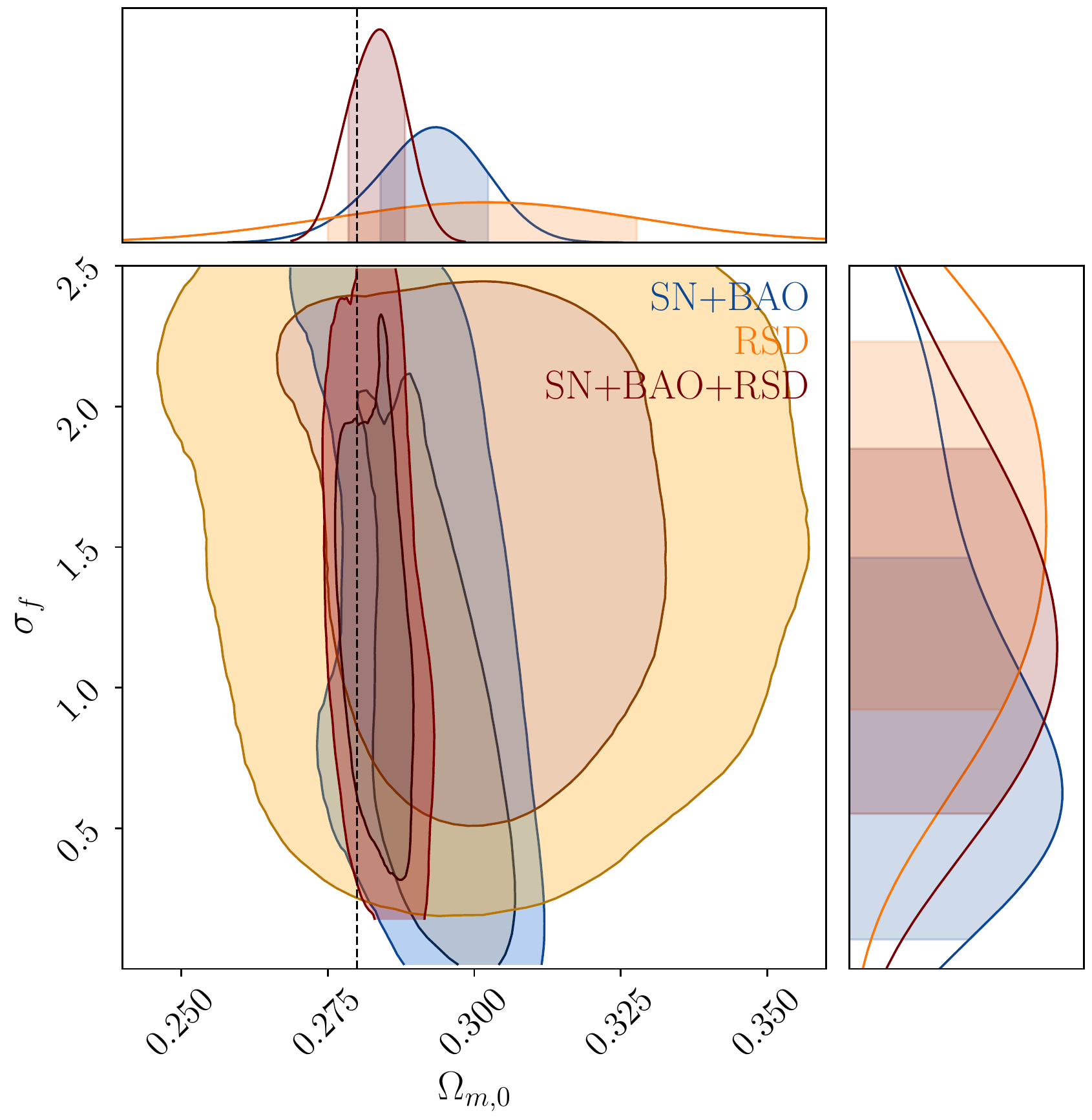}
    \caption{Posteriors of the kernel hyperparameter $\sigma_f$ vs. $\Omo$ from various chains---assuming perfect knowledge of \seighto. It is seen that in all cases, those samples of $f_{\rm DE}$ with a value of $\Omega_{\rm m,0}=0.28$ (corresponding to the truth) require a deviation from the mean (\emph{i.e.} $\sigma_f\neq0$). Furthermore, ``background-only''  probes (SN \& BAO) are consistent with no deviations from the mean (\emph{i.e.} $\sigma_f=0$) provided $\Omega_{\rm m,0}\sim0.31$ - close to \lcdm's best-fit value. However, RSD are more sensitive to the evolution of $\fde$ and do not allow for $\fde=1$ (\emph{i.e.} $\sigma_f\neq0$), though the value of $\Omo$ is essentially unconstrained.}
    \label{fig:posteriors_hyper_col}
\end{figure}
\subsection{\tpdf{($\Omega_{\rm m,0}$,\,\seighto)} known}

The left panel of Fig.~\ref{fig:h_om_s8} shows the reconstructed \fde\ assuming perfect knowledge of $\Omo$ and $\seighto$ using various combinations of SNIa, BAO and RSD data. 
The lower panels throughout this paper show the percentage error with respect to the fiducial cosmology used to generate the data---for a given quantity X, we define $\frac{\Delta X}{X}\equiv \left(\frac{X-X^{\rm fid}}{X^{\rm fid}}\right)\times100$.
The DE evolution \fde\ is well reconstructed at low-$z$ where the data are present. 
At high-$z$, our choice of mean function $\bar{f}_{\rm DE}=1$ ensures a EdS behaviour as expected. However, at intermediate  $1.5\lesssim z \ll 20$, \fde\ is not reconstructed well. 
This is understandable because  this redshift range is in the matter-dominated era, where the constraining power on \fde\ becomes negligible. 
However, because of the knowledge of the fiducial \Omo, the expansion history is reconstructed relatively well as shown on the right-hand panel. 
\begin{figure*}
    \centering
    \includegraphics[width=\textwidth]{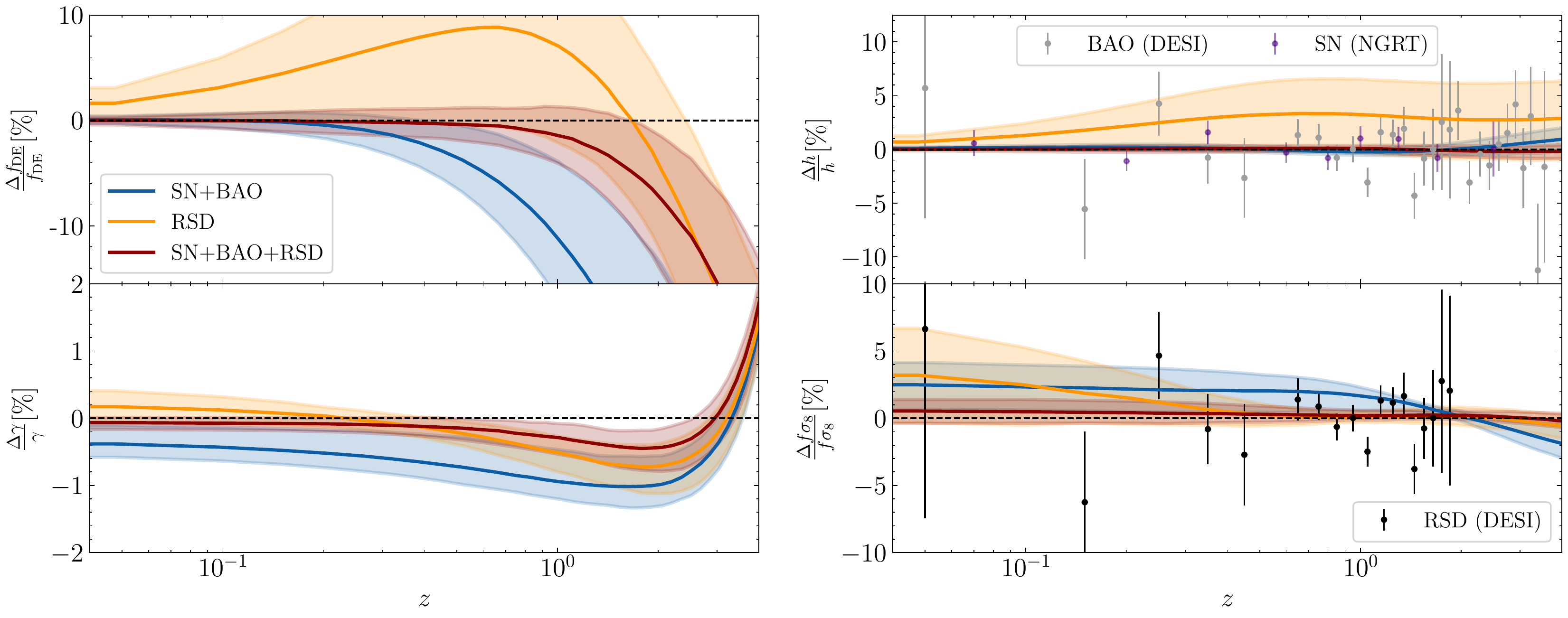}
    \caption{Upper panels: Percent errors in the reconstructed background quantities \fde\ (left) and $h(z)$ (right) assuming perfect knowledge of \seighto, marginalized over \Omo. Lower panels: Reconstructed (perturbed) quantities $\gamma\equiv\ln{f}/\,\ln{\Omega_{\rm m}}$ and $f\sigma_8$ from joint analyses. Solid lines and shaded regions show the median and $1\sigma$ confidence regions around it, respectively. As shown in the upper panels, when \Omo\ is free to vary, \fde\ is not well reconstructed with SN+BAO, as background-only probes can compensate for the ``lack of DE'' at large-$z$ with larger values of \Omo. This bias in \Omo\ is reflected in the \fseight\ reconstructions shown in the bottom-right panel, being marginally consistent with the fiducial solution. Adding data from RSD (in red) yields a better reconstruction of the quantities of interest.
    } 
    \label{fig:residuals_om_free}
\end{figure*} 
\noindent In this particular choice of fiducial cosmology, \fde\ grows with redshift, and thus contributes more to the expansion rate of the universe at large-$z$, compared to \lcdm.
Our choice of mean function $\bar{f}_{\rm DE}=1$ induces a slight bias towards \lcdm, as in the absence of data \fde\ will go back to its mean, which explains why $h(z)$ is underestimated at intermediate redshifts. 
At higher-redshifts, this bias becomes negligible and an accurate estimation of \Omo\ is more relevant, since $h(z\gg1)\sim\left[\Omo(1+z)^3\right]^{1/2}$. 
However, this bias induced in $h(1.5<z<25)$ is below $2\%$. 

\subsection{\tpdf{\seighto\ } known}
We now study the effect of releasing the assumption of perfect knowledge of $\Omo$.
Instead, we now vary \Omo\ together with the hyperparameters in the MCMC runs. 
The reconstructions are shown in Fig.~\ref{fig:residuals_om_free}.
Without perfect knowledge of \Omo, the uncertainties on the reconstructions increase, as expected. 
The SN+BAO reconstructions underestimate \fde\, while the RSD-only reconstruction overestimate it for $z\leq 1$. 
Combining the three probes results in a more accurate and precise reconstruction of \fde\ on this redshift range.
Interestingly, while the
SN+BAO reconstructions of $h$ are overestimated at high-redshift, and the RSD-only reconstructions overestimate $h$ over the whole redshift range, the reconstruction of $h$ with the three probes is accurate and precise over the whole range of redshifts. We also note that the relative errors for $f\sigma_8(z)$ measurements are larger than those of BAO and SNIa at low-$z$,  which further explains the larger uncertainty at low-$z$.

\begin{figure*}
    \centering
    \includegraphics[width=\textwidth]{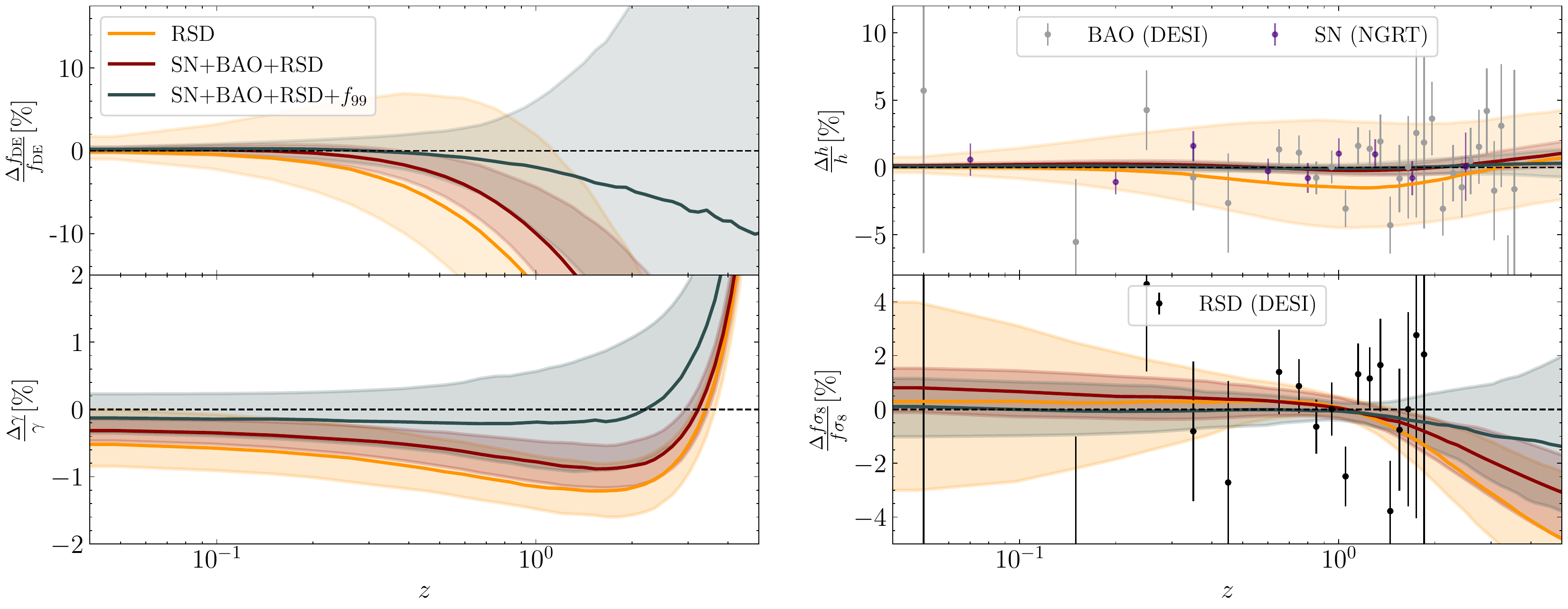}
    \caption{Residuals from a joint analysis, marginalized over \Omo\ and \seighto, assuming GR. We also compare the reconstructions with and without $f_{99}\equiv f_{\rm DE}(z=99)$---\emph{c.f.} Section~\ref{sec:bias}. Although the uncertainties increase significantly, the reconstructions including $f_{99}$ are in $1\sigma$ agreement with the fiducial solutions throughout the redshift range $0<z\lesssim10^2$.}
    \label{fig:residuals_om_s8_free}
\end{figure*}

\smallbreak 

In the lower panels of Fig. \ref{fig:residuals_om_free}, we show the reconstructions of the growth index $\gamma(z)={\ln f}/{\ln \Omm}$ and $\fseight(z)$ when marginalizing over \Omo, assuming GR. 
The dashed line shows the fiducial solution, for which $\gamma \simeq 0.55$, with a slight redshift evolution---see \cite{Polarski:2016ieb,Calderon:2019jem,Calderon:2019vog}. 
As explained before, the background probes (SN+BAO) are not able to constrain the value of \Omo\ (and thus to accurately reconstruct \fde---see upper panels of Fig.\ref{fig:residuals_om_free}) which in turn is reflected in the reconstructed growth history \fseight, particularly at low-$z$. When including RSD, and assuming the fiducial value of \seighto, the reconstructions (in orange) are much more accurate, both in $\gamma$ and \fseight. 
From the well-known relation $f\simeq\Omm^\gamma$, we see that larger values of \Omo\ will require lower values of $\gamma$ in order to keep the agreement with observations, as reflected in the bottom-left panel of Fig.\ref{fig:residuals_om_free} for the case of SN+BAO. %A joint analysis allows for an accurate enough determination of \Omo\ provided \seighto~ is known, as can be seen on the right panel of Fig~\ref{fig:posteriors_hyper}.% , where we show the value of $\Omo$ as a function of the GP hyperparameters for XXX samples of $fde$. 

The bias in the background-only (SN+BAO) reconstructions is understood by looking at the posterior distributions of $(\sigma_f,\Omo)$, shown in Fig. \ref{fig:posteriors_hyper_col}. 
Each point in parameter space correspond to a given realization of \fde\, and expansion history characterized by certain combinations of $(\sigma_f,\ell_f)$ and \Omo---while $\seighto$ is fixed to its fiducial value. 
For the background-only probes (SN+BAO), the data are consistent with no deviations from the mean ($\sigma_f=0$), provided a high value of \Omo$\sim0.31$, while those samples with the true (fiducial) value of \Omo$=0.28$ require $\sigma_f\neq0$. However, redshift-space distortions are more sensitive to the nature of DE, and thus do not allow for a \lcdm-like evolution ($\fde=1$) and require $\sigma_f\neq0$ - at the price of losing the constraining power on \Omo. As mentioned before, when combining the three different datasets, these effects compensate and we get an accurate reconstruction of \fde, and a clearer detection of $\sigma_f\neq0$ (\emph{i.e.} a deviation from the mean---\lcdm), by narrowing down the allowed range of \Omo.

\subsection{\tpdf{($\Omega_{\rm m,0}$,\,\seighto)} unknown}\label{section:om_s8_free}
 
We now release the assumption of perfect knowledge of $\Omo$ and \seighto\ and marginalize over them. 
Fig.~\ref{fig:residuals_om_s8_free} shows the relative errors in the reconstructions of \fde\ and $h$ after marginalizing over the two cosmological  parameters and the two hyperparameters.
As previously, \fde\ cannot be reconstructed accurately at $z\geq 1$. 
The RSD-only reconstruction underestimate $\fde$ from low redshift, and the uncertainties are quite large. 
The smaller uncertainties in the combined probes case yields a reconstruction within $2\sigma$ of the expansion history, with a slight overestimation at high-redshifts. This stems from the fact that we are biased towards \lcdm's best-fit value of \Omo$\sim0.31$. Furthermore, since $\sigma_{8,0}$ is now a free parameter, and is degenerate with \Omo, we loose the ability to ``narrow down'' \Omo\ to its ``true'' (fiducial) value and thus, to accurately reconstruct \fde. This bias will be reflected on the growth evolution \fseight, which depends not only on $h(z)$ but also on $\Omm(z)$ via the source term for the perturbations $\delta$- see lower panels of Fig. \ref{fig:residuals_om_s8_free}, particularly the left panel showing $\gamma(z)$.

The lower panels of Fig. \ref{fig:residuals_om_s8_free} also show  the reconstructed perturbed quantities when marginalizing over both \Omo\ and \seighto. In this case, as understood from the background reconstructions in the upper panel, the degeneracy between \Omo\ and \seighto\ causes the reconstructions of \fde\ to be less accurate, even in the joint SN+BAO+RSD case. From the $h(z)$-data standpoint, having ``less'' Dark Energy ($\fde<\fde^{\rm fid}$) at large-$z$ can be compensated by having more matter $\Omo>\Omo^{\rm fid}$. Which enforces $\seighto<\seighto^{\rm fid}$,  while keeping the agreement with RSD observations. Thus, when $\seighto$ is also free to vary, we lose the constraining power on $\Omo$ provided by RSD measurements and therefore we do not reconstruct $\fde(z\geq0.6)$. The $2\%$ percent bias induced in $h(z\geq2)$ translates into a bias in $\fseight(z\geq2)$ as can be seen in the lower-right panel of Fig. \ref{fig:residuals_om_s8_free}. Despite the $2\sigma$ agreement with both $h(z)$ and $\fseight(z)$ measurements, the bias in the values of \Omo\ and \seighto\ are clearly seen in the evolution of $\gamma(z)$ --- shown on the left panels of Fig.  \ref{fig:residuals_om_s8_free}.

\begin{figure}[b]
    \centering
    \includegraphics[width=\columnwidth]{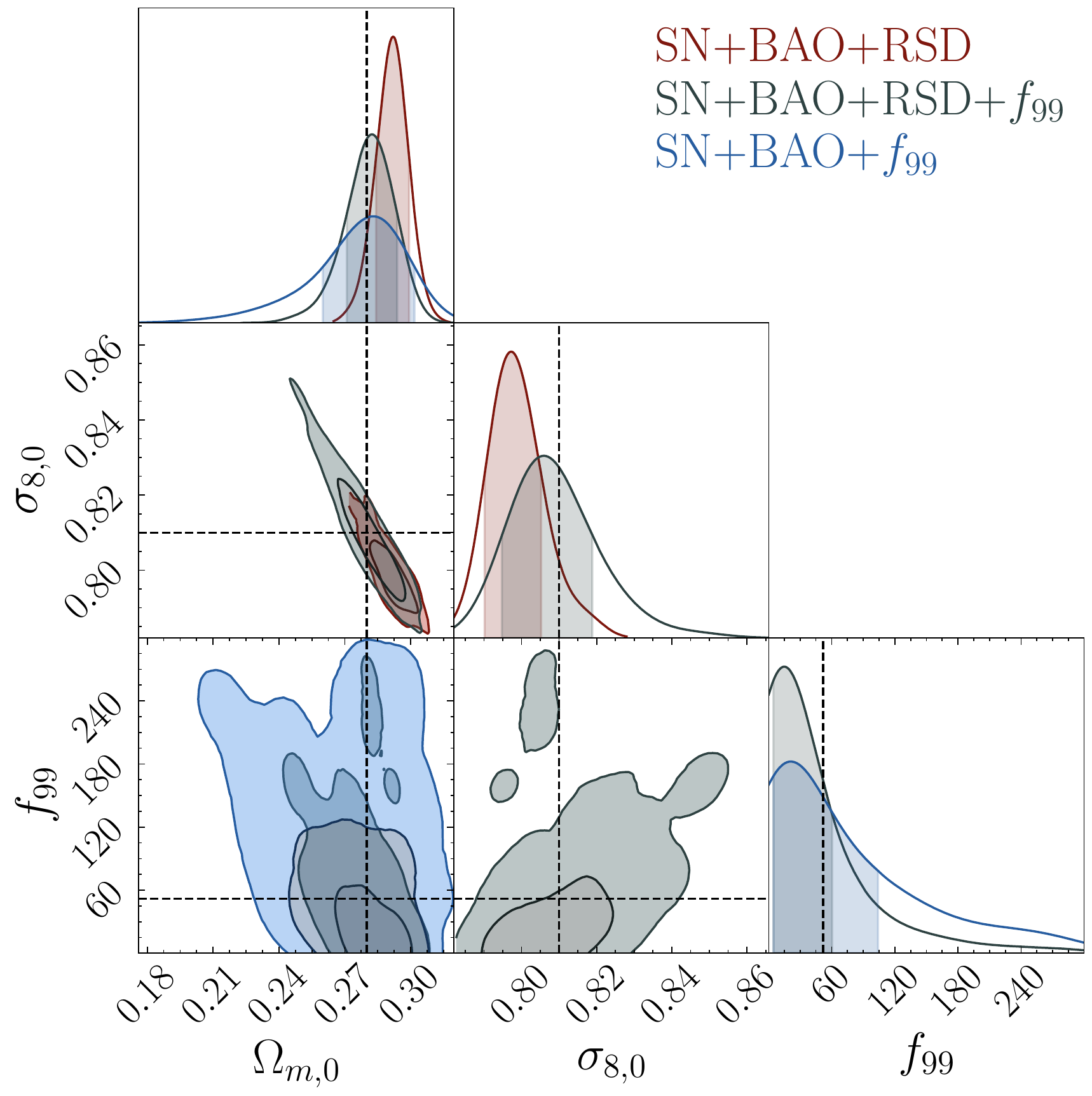}
    \caption{Posterior distributions for the cosmological parameters in a combined analysis.
    By allowing for a non-negligible contribution from DE at large-$z$, encoded in $f_{99}$, we remove the bias seen before. Dashed lines indicate the fiducial values.% used to generate the data.
    }
    \label{fig:posteriors_unbiased}
\end{figure}

%%%%%%%% essai: 
\subsection{Removing the Biases}
\label{sec:bias}

As discussed before, assuming \lcdm\ (\emph{i.e. $\fde=1$}) at high-$z$ is a reasonable assumption but can induce a bias in the reconstructions if the contribution from DE is non-negligible at such redshifts. As can be seen from Fig. \ref{fig:posteriors_unbiased}, our contours are biased and shifted towards \lcdm's best-fit value - \emph{i.e.} a higher $\Omo$ and lower \seighto\ to compensate for the lack of DE. To deal with such a bias, we introduce an additional degree of freedom in the MCMC runs. Namely, $f_{99}\equiv\fde(z=99)$ characterizing the DE density at $z\sim10^2$, when we start our integration. More specifically, for each step in the MCMC, we impose a value for \fde\ at both ends of the redshift range, namely $\fde=1$ at $z=0$ and $\fde=f_{99}$ at $z=99$---while $f_{99}$ is free to vary in the range $f_{99}\in[0.01,300]$. The behaviour of $\fde$ between $0<z<99$ is solely determined by the data at low-redshift, without biasing our parameter inferences by assuming $\Lambda$ in the past---particularly those of \Omo\ and \seighto.
This enables us to cover a wide variety of DE models, including those where the DE density grows with redshift and contributes as much as $\sim0.1\%$ of the matter density at $z=10^2$. The upper bound in our prior is motivated by the fact that observations require an epoch of matter domination for structures to efficiently form at early times ($\delta\propto a(t)$), and the latest \texttt{Pantheon+} analysis \cite{Brout2022vxf} on CPL cosmologies constrain $-1.1\lesssim w_0\lesssim -0.6$ and $-1.3\lesssim w_a\lesssim 0.2 $. 
We note that for our choice of fiducial cosmology (\emph{i.e. $w_0=-0.95,w_a=0.3$}), the normalized energy density of DE is $\fde(z=99)\simeq52$.

\begin{figure*}
    % \centering
    \includegraphics[width=\textwidth]{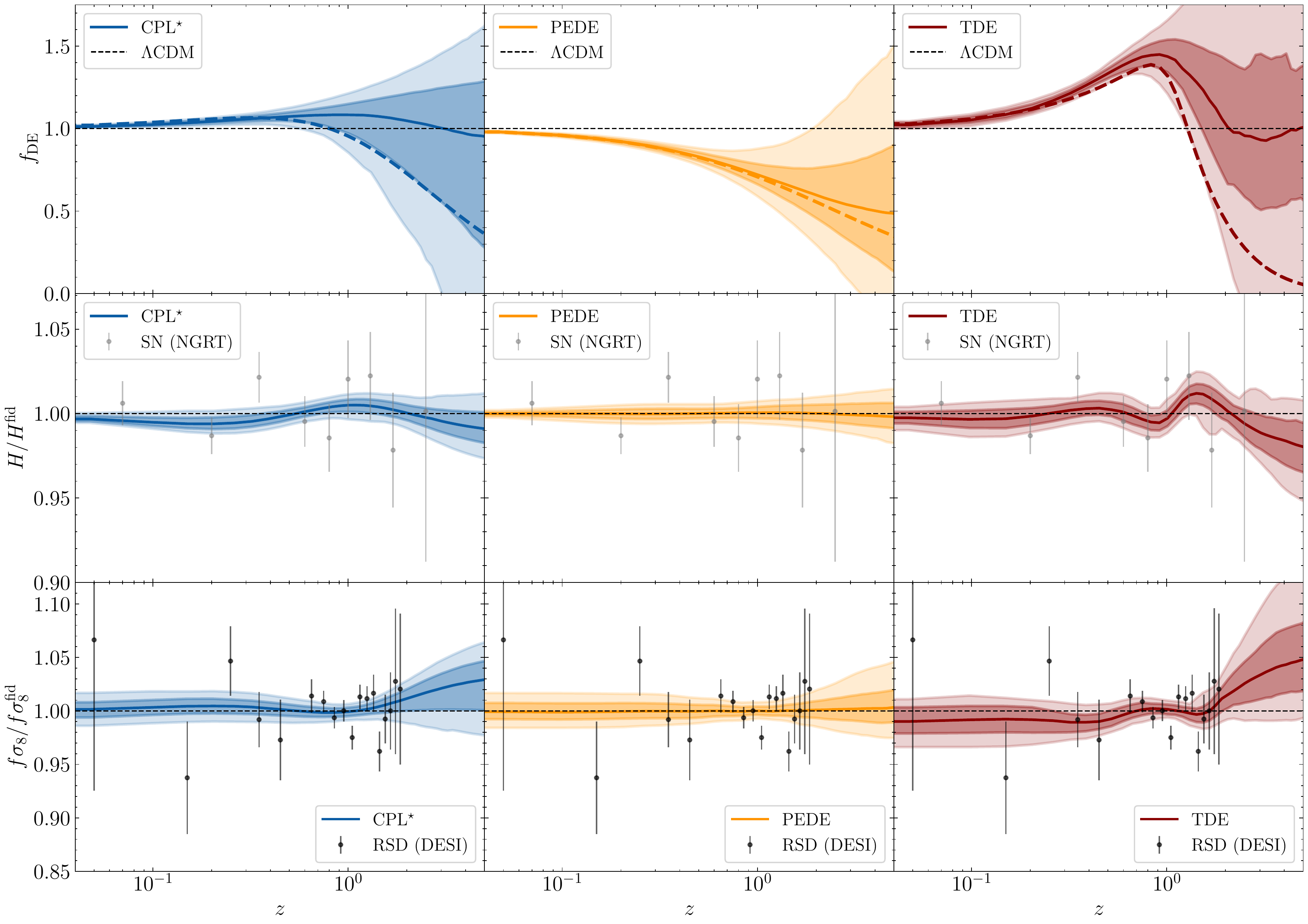}
    \caption{Reconstructions of various (different) fiducial cosmologies.
    Solid lines and shaded regions correspond to median and $95\%$ C.L. around it, respectively. Dashed-colored lines in the upper panels depict the ``true'' behaviour of $\fde(z)$ for the different cosmologies. In the lower panels, the corresponding observables $H(z)$ and $\fseight(z)$ normalized to the correct solution in the fiducial cosmology.}
    \label{fig:fiducial_cosmos}
\end{figure*}

\begin{table}[t]
    \centering
    \label{tab:param_constraints}
    \begin{tabular}{cccc}
        \hline
		Datasets & $\Omega_{m,0}$ & $\sigma_{8,0}$ & $f_{99}$ \\ 
		\hline
		SN+BAO+RSD & $ 0.2920^{+0.0073}_{-0.0077}$ & $ 0.7972^{+0.008}_{-0.007}$ &---\\ SN+BAO+$f_{99}$ & $0.283^{+0.019}_{-0.023}$ & --- & $21^{+83}_{-17}$ \\ 
		SN+BAO+RSD+$f_{99}$ & $0.282^{+0.011}_{-0.012}$ & $0.806^{+0.013}_{-0.011}$ & $15^{+45}_{-11}$ \\ 
		\hline
    \end{tabular}
    \caption{Forecasted constraints on cosmological parameters from  joint analyses using data from Stage-IV surveys. We remind the reader that the fiducial values used to generate the data are $\Omo^{\rm fid}=0.28$, $\seighto^{\rm fid}=0.81$ and $f_{99}^{\rm fid}=51.64$.}
\end{table}

\smallbreak
In Fig. \ref{fig:residuals_om_s8_free} we compare the reconstructions with and without $f_{99}$. The biased case (in dark-red) is in $1\sigma$ agreement with the background observable $h(z)$ and within $2\sigma$ agreement with the growth observable $\fseight(z)$, despite the clear discrepancy with the true DE evolution at $z\gtrsim 1$, depicted in the left panels. %As mentioned before, the higher $\Omo$ needed to keep the agreement with $H(z)$-data is compensated by lower $\seighto$ to keep the agreement with RSD data, as seen in the posteriors shown in Fig. \ref{fig:posteriors_unbiased}. This bias in the ($\Omo,\seighto$)-plane is also seen in the reconstructions of $\gamma(z)$.
The dark gray line in Fig. \ref{fig:residuals_om_s8_free} shows the reconstructions when including the $f_{99}$ parameter in the runs, increasing the uncertainty on the reconstructions but being consistent with $\fde^{\rm fid}$ throughout the redshift range.

\section{Reconstructing various Fiducial Cosmologies}
\label{sec:varcosmo}

In this section, we apply our methodology to reconstruct other (different) cosmologies. In particular, we focus on cosmologies where the behaviour of \fde\ is different than the one presented in the previous section of this paper. First, we reconstruct a CPL model with the best-fit parameters as obtained in the latest joint
\texttt{Pantheon+} analysis \cite{Brout2022vxf} (including BAO and Planck data).
We also consider two additional examples, namely the Phenomenological Emergent Dark Energy (PEDE) \cite{LS2019} and the Transitional Dark Energy (TDE) \cite{Keeley2019} models. 

\begin{itemize}
    \item $\rm CPL^{\rm \star}$: We take the best-fit values for the CPL cosmology in the latest \texttt{Pantheon+} analysis \cite{Brout2022vxf}. These correspond to $\theta^{\rm fid}=\{\Omo^{\rm fid}=0.31,w_0^{\rm fid}=-0.85,w_a^{\rm fid}=-0.6\}$ and we fix $\seighto^{\rm fid}=0.81$. We denote this $\rm CPL^{\rm \star}$ to distinguish it from the fiducial cosmology presented in \S\ref{sec:valid}.
    \item PEDE: Similarly, for the PEDE model we fix $\theta^{\rm fid}=\{\Omo^{\rm fid}=0.31,\seighto^{\rm fid}=0.81\}$ with $w(z)$ given by Eq. (6) in \cite{LS2019}
    \item TDE: For the TDE model \cite{Keeley2019}, the cosmological parameters are $\theta^{\rm fid}=\{\Omo^{\rm fid}=0.31,w_0^{\rm fid}=-0.8,w_1^{\rm fid}=-2.0,z_t^{\rm fid}=1.0,\Delta_z^{\rm fid}=0.2,\seighto^{\rm fid}=0.81\}$ with the expression for $w(z)$ given by Eq. (4.2) in \cite{koo2021bayesian}. 
\end{itemize}

The $\fde(z)$ and $h(z)$ reconstructions for the corresponding cosmologies are shown in Fig.\ref{fig:fiducial_cosmos}, when marginalizing over the cosmological+hyper parameters and using SN+BAO+RSD data.

The $\fde$ (and thus $h(z)$) reconstructions in the CPL and PEDE cosmologies are accurate within $1\sigma$ across the entire redshift range $0<z<4$. %, even without including the additional parameter $f_{99}$.
In such cosmologies, the energy density of DE smoothly decreases with redshift, and as such, contribute even less than the $\Lambda$-term to the energy budget of the universe at $z\sim10^2$, and thus our assumption of an EdS universe at such redshift is well-justified.
In the case of larger departures from \lcdm, as in the case of TDE shown in red, the reconstructions are able to capture the ``increasing'' trend in $\fde$ at low-$z$ ($z\lesssim1$), where the data are abundant and DE dominates, but tend to go back to the mean ($\fde\sim1$) where the constraining power on DE is lost---both because the quality of the data decreases, and because matter becomes the dominant component. This translates into $\lesssim2\%$ deviations in terms of $h(z)$, while being consistent within $2\sigma$ with the fiducial expansion and growth histories.
Interestingly, in all cases the $\fde$ reconstructions are in tension---at $\gtrsim2\sigma$---with \lcdm~ ($\fde\equiv 1$) for $z\lesssim1$.
 
\section{Discussions and Conclusions}

\label{sec:ccl}

In this work a method is presented for the joint reconstruction of the cosmic expansion and growth histories up to 
large redshifts $z\lesssim 100$ using Gaussian processes. 
When using Gaussian processes, the problem to be solved is the scarcity or even the total absence of data for $z\gtrsim 2$. 
On the other hand as we expect most DE models to behave like a quasi Einstein de Sitter universe on these intermediate redshifts, we chose  to model $\fde\sim\mathcal{GP}$ with mean function 1 while only assuming flat FLRW+GR. 
This is a sufficient condition to impose an EdS behaviour at high-redshifts.
These reasonable assumptions leave the DE evolution to be solely determined by the data at low-$z$.
Hence we emphasize the flexibility of our method with respect to the possible DE phenomenologies at low-$z$ where DE dominates and where specific models leave their imprint in observations. 
We tested the efficiency of our method in jointly reconstructing the cosmic expansion and growth histories by simulating mock data for the upcoming generation of cosmological surveys and for various dark energy models. 
In \S\ref{sec:valid} we extensively discuss the reconstruction of a fiducial CPL cosmology with $w_0=-0.95$ and $w_a=0.3$ (hence with a rather strong evolution of $w$ upwards for growing $z$) which is at the limit of being excluded by current data (\emph{e.g.} \texttt{Pantheon+} \cite{Brout2022vxf}), in other words we choose to test our method with a ``worst-case'' scenario which departs substantially from a cosmological constant at large redshifts while still obeying an EdS like behaviour at high-$z$. The successful reconstruction of this fiducial cosmology implemented through the introduction of the parameter $f_{99}$ means that it will work as well, perhaps even without making use of $f_{99}$, for most DE models including Phenomenological Emergent Dark Energy (PEDE) or Transitional Dark Energy (TDE) models (see Fig. \ref{fig:fiducial_cosmos}) which generically do not contribute much to the energy budget at high-$z$. 
We have shown that $f_{\rm DE}$ reconstruction from stage-IV surveys can potentially detect deviations from \lcdm\ at more than $2\sigma$. 
We believe the joint reconstruction method presented here, which was refined in order to capture a small amount of DE at high redshifts when $\Omega_{\mathrm{DE}}\to 0$ though not as quickly as in $\Lambda$CDM, can address such tracking (non-interacting smooth) DE as well. %though this is left for future work. 
We conjecture that by including information at higher-$z$ (\emph{e.g.} CMB distance priors), we could potentially ``constrain'' the contribution of (early) DE at $z\sim z_{\rm rec}$ by further constraining $f_{99}$. 
We also expect that our approach can be extended to more cosmologies, and that a non-vanishing spatial curvature or the effect of massive neutrinos can be constrained accurately. We finally expect interesting results can be obtained for modified gravity DE models and we leave all these developments for future work.

\section*{Acknowledgements}
BL acknowledges the support of the National Research Foundation of Korea (NRF-2019R1I1A1A01063740) and the support of the Korea Institute for Advanced Study (KIAS) grant funded by the government of Korea. 
AS would like to acknowledge the support by National Research Foundation of Korea NRF2021M3F7A1082053, and the support of the Korea Institute for Advanced Study (KIAS) grant funded by the government of Korea.
AAS was partly supported by the project number 0033-2019-0005 of the
Russian Ministry of Science and Higher Education.

%%%%%%%%%%%%%%%%%%%%%%%%%%%%%%%%%%%%%%%%%%%%%%%%%%

% %%%%%%%%%%%%%%%%% APPENDICES %%%%%%%%%%%%%%%%%%%%%
% % \clearpage
% \appendix

% \nocite{*}

\bibliography{_biblio}% Produces the bibliography via BibTeX.

\end{document}